\documentclass[a4paper,12pt]{article}
\title{Kapitza pendulum effect in a weakly disordered amorphous  magnet}
\author {I.A.Fomin\\
P. L. Kapitza Institute for Physical Problems, \\
ul. Kosygina 2, 119334 Moscow,Russia}
\date{ }
\begin{document}

\maketitle
\begin{abstract}
Effect of a "random anisotropy" type disorder on orientation of
magnetization in an amorphous magnet is considered. It is shown,
that principal corrections to the free energy of the magnet
originate from the fluctuation of the order parameter in the
directions of its degeneracy. The types of disorder are found,
which lift the continuous degeneracy and suppress the
Larkin-Imry-Ma mechanism of disruption of the long-range order in
continuously degenerate systems.


\end{abstract}
 {\bf 1.} Continuously degenerate ordered systems react strongly
 even on a weak disorder. It has been demonstrated by Larkin \cite{Lark}
 in connection with the translationally degenerate system and by
 Imry and Ma \cite{imry} for systems with rotational degeneracy that in the
 spaces with dimensionality $d\leq 4$ an arbitrary weak disorder
 disrupts the long-range order. Let magnetization $\mathbf{M}$ be the order
 parameter, $\xi$- the correlation length and $g$--a small
 parameter, characterizing the weakness of the disorder
 (it will be specified later). According to the Imry and Ma argument
  orientation of $\mathbf{M}$ can vary
 in space to adjust to the fluctuations of the disorder on a scale
$L_{\eta}\sim \xi/g$. This nonuniform state has lower energy then
 the uniform for a relative value of the order of $g^2$. It
is assumed in this argument that the continuous degeneracy is not
lifted by corrections to the energy of a lower order in $g$. The
known examples \cite{aharon,minchau,wehr} demonstrate that some
types of disorder can orient $\mathbf{M}$, i.e. to lift a
continuous degeneracy. It was demonstrated also \cite{fom1} that
for the superfluid $^3$He in aerogel corrections of the lowest
order in $g$ can suppress the Larkin-Imry-Ma effect.

In a present paper the argument of Ref.\cite{fom1} is applied to a
more simple object -- amorphous ferromagnet with the disorder of
the "random anisotropy" type  \cite{plish} in a vicinity of the
Curie temperature $T_c$. For such ferromagnet the particular types
of disorder are found, which lift the continuous degeneracy and
suppress the effect of Larkin-Imry-Ma. As an example the
three-dimensional ferromagnet is considered with the weak random
anisotropy in in the $x,y$-plane  \footnote{the example is
suggested by E. I. Kats}. It is shown that such disorder orients
magnetization in $z$-direction. The orienting effect  manifests
itself in a region where the disorder is still small and can be
treated  as a perturbation \cite{LarkOv}.

{\bf 2.} The Landau expansion of the free energy of an amorphous
ferromagnet in the powers of $\mathbf{M}$ in a vicinity of $T_c$
has the form:
$$
 F_f=F_n+\int d^3r[\tau
 M_jM_j+\frac{b}{2}(M_jM_j)^2+\xi_0^2\frac{\partial M_j}{\partial x_n}
 \frac{\partial M_j}{\partial x_n}+\eta_{jl}({\bf
 r})M_jM_l+\kappa_{jl} M_jM_l].  \eqno(1)
$$
The coefficients here have the usual meaning: $\tau=(T-T_c)/T_c$;
$b=const., b>0$, the spatial rigidity is denoted as
 $\xi_0^2$. The units for  $\bf{M}$ are chosen to yield the proper
 dimensionality to the integrand. Random anisotropy is
 described by the term $\eta_{jl}({\bf r})M_jM_l$, where $\eta_{jl}({\bf r})$-
 a real symmetric random tensor, its trace $\eta_{nn}({\bf r})/3$
 describes the local shift of the transition temperature and the
 remaining part $\eta_{jl}^{(a)}({\bf r})=\eta_{jl}({\bf
r})-\delta_{jl}(\eta_{nn}({\bf r})/3)$ -- the local splitting of
 $T_c$ for different components of $\mathbf{M}$. The ensemble of
 tensors $\eta_{jl}({\bf r})$ is assumed to be spatially uniform on the average.
 The average anisotropy $\kappa_{jl}$ is written as a separate
 term, i.e. $<\eta_{jl}({\bf r})>=0$. The results will be expressed in terms of
 the Fourier transforms of binary correlation functions  $\eta_{jl}({\bf r})$
 in a limit  $k\to 0$:\quad $\Phi_{jlmn}(0)=
 <\eta_{jl}(\textbf{k})\eta_{mn}(-\textbf{k})>\mid_{k=0}$. For the
 moment no further assumptions are made about these functions.
The equilibrium $\mathbf{M}({\bf r})$ is found from the equation:
$$
\tau M_j+\kappa_{jl}M_l+\eta_{jl}({\bf
r})M_l-\xi_0^2\frac{\partial^2 M_j}{\partial x_n^2}+bM^2M_j=0.
\eqno(2)
$$
To separate the effect of the disorder on  orientation of the
order parameter the case $\kappa_{jl}=0$ is considered first.
Treating $\eta_{jl}({\bf r})$ as a small perturbation we assume
that the solution of Eq.(2) has a form: $M_j({\bf
r})=\overline{M}_j+m_j({\bf r})$, where $m_j({\bf r})$ is a small
{\it fluctuation}:
 $\mid m_j({\bf r})\mid\sim\mid\eta\mid^{\gamma}\overline{M}$, $\gamma>0$, its
 average over the volume with a linear dimensions smaller then
 $L_{\eta}$ vanishes:  $<m_j(\textbf{r})>=0$.
The expansion of Eq. (2) over $\eta_{jl}$ and $m_j$  and
separation of smoothly and fast varying terms renders equations
for
 $m_j({\bf r})$ and $\overline{M}_j$.
 Possible dependence of $\overline{M}_j$ on $\textbf{r}$ would effect
 terms of the higher order in $\eta_{jl}$ then considered here, so we can
 assume $\overline{M}_j=const.$. For brevity the bar over
$\overline{M}_j$ in what follows is dropped.
$$
\tau M_j+b[M^2M_j+2<m_jm_l>M_l+<m_lm_l>M_j]+<\eta_{jl}m_l>=0,
\eqno(3)
$$
$$
\tau m_j+b[M^2m_j+2M_jM_lm_l+(m_lm_l)m_j]-\xi_0^2\frac{\partial^2
m_j}{\partial x_n^2}=-\eta_{jl}M_l. \eqno(4)
$$
The local fluctuations of magnetization $m_j(\textbf{r})$ in
directions parallel and perpendicular to $\bf{M}$ have to be
treated differently. Projection of Eq. (4) on the unit vector
$\hat{\mu}=\mathbf{M}/M$ renders:
$$
\tau m_{\mu}+3bM^2m_{\mu}-\xi_0^2\frac{\partial^2m_{\mu}}{\partial
x_n^2}=-\eta_{\mu\mu}M, \eqno(5)
$$
where $m_{\mu}=\hat\mu_lm_l$ and
$\eta_{\mu\mu}\equiv\eta_{jl}\hat{\mu}_j\hat{\mu}_l$. Here and in
what follows the summation over the repeated  Greek  indices is
not assumed.

The cubic in $m_l$ term can be omitted here. The equation is
linear and can be solved by Fourier transformation:
$$
m_{\mu}(\textbf{k})=-\frac{\eta_{\mu\mu}(\textbf{k})}{2|\tau|+
\xi_0^2k^2}M, \eqno(6)
$$
Then for the average fluctuation we have:
$$
<m_{\mu}(\textbf{r})m_{\mu}(\textbf{r})>=
M^2\int\frac{<\eta_{\mu\mu}(\textbf{k})\eta_{\mu\mu}(\textbf{-k})>}{[2|\tau|+
\xi_0^2 k^2]^2}\frac{d^3k}{(2\pi)^3}. \eqno(7)
$$
The integral here converges at  $k^2\sim |\tau|/\xi_0^2\sim
1/[\xi(T)]^2$. For that region we assume
$\Phi_{\mu\mu\mu\mu}(\textbf{k})\approx\Phi_{\mu\mu\mu\mu}(\textbf{0})$,
then
$$
<m_{\mu}({\bf r})m_{\mu}({\bf r})>=\frac{1}{8\pi}
\frac{\Phi_{\mu\mu\mu\mu}(0)} {\xi_0^3}\frac{M^2}{\sqrt{2|\tau|}},
\eqno(8)
$$
For the validity of the above argument the average $<m_{\mu}({\bf
r})m_{\mu}({\bf r})>$ has to be small in comparison with $M^2$, or
$g\equiv\frac{1}{8\pi}
\frac{\Phi_{\mu\mu\mu\mu}(0)}{\xi_0^3\sqrt{2|\tau|}}\ll 1 $. The
introduced here $g$ serves as a small parameter of the present
theory.

To find the transverse with respect to $\textbf{M}$ components of
 $m_j(\textbf{r})$ the projection of Eq. (4)  on the unit vectors
 $\hat{\bf\lambda}$ and $\hat{\bf\nu}$, forming together with $\hat{\bf\mu}$
 orthogonal basis are used:
$$
\tau m_{\lambda}+b[M^2m_{\lambda}+(m_lm_l)m_{\lambda}]+
\eta_{\lambda\mu}M -\xi_0^2\frac{\partial^2m_{\lambda}}{\partial
x_n^2}=0, \eqno(9)
$$
and an analogous equation obtained by substitution of $\lambda$ by
$\nu$. If the anharmonic  term here is dropped  the average square
of the transverse fluctuation would be proportional to the
diverging integral:
$$
<m_{\lambda}(\textbf{r})m_{\lambda}(\textbf{r})>=
M^2\int\frac{<\eta_{\lambda\mu}(\textbf{k})\eta_{\lambda\mu}(\textbf{-k})>}
{[\xi_0^2k^2]^2}\frac{d^3k}{(2\pi)^3}. \eqno(10)
$$
The same applies for $<m_{\nu}(\textbf{r})m_{\nu}(\textbf{r})>$.
To make the expressions for the transverse fluctuations finite the
anharmonic terms have to be taken into account. That can be done
in the mean field approximation. In Eq. (9) the substitution is
made:
$$
(m_lm_l)m_{\lambda}\approx<m_lm_l>m_{\lambda}+2<m_{\lambda}m_l>m_l.
\eqno(11)
$$
The sum $\tau+bM^2$ also has to be expressed in terms of $m_j$ and
$\eta_{jl}$ with the accuracy up to the second order:
$$
\tau+bM^2=-\frac{1}{M}<\eta_{\mu
l}m_l>-2b<m_{\mu}m_{\mu}>-b<m_lm_l>.  \eqno(12)
$$
  The average $<\eta_{\mu l}m_l>$  is expressed as the integral,
  diverging at large wave vectors $k$. The diverging part can be
  included in $\tau$ and $\kappa_{jl}$ as it was done in Ref.\cite{LarkOv}.
  The remaining part of the integral is not singular at small
 $k$. Keeping only singular terms we arrive at:
$$
\tau+b(1+\sigma_{\lambda\lambda}+\sigma_{\nu\nu})M^2=0, \eqno(13)
$$
where $M^2\sigma_{\lambda\lambda}\equiv<m_{\lambda}m_{\lambda}>$
and $M^2\sigma_{\nu\nu}\equiv<m_{\nu}m_{\nu}>$. The remaining
freedom in the orientation of the vectors
 $\hat{\bf\lambda}$ and $\hat{\bf\nu}$ can be used for turning to
 zero the off-diagonal component $\sigma_{\lambda\nu}$. Then Eq.
  (9) reads as:
$$
2bM^2\sigma_{\lambda\lambda}m_{\lambda}-
\xi_0^2\frac{\partial^2m_{\lambda}}{\partial
x_n^2}=-M\eta_{\lambda\mu}. \eqno(14)
$$
An analogous equation can be written for $m_{\nu}$. Eq. (14) has
solution:
$$
m_{\lambda}({\bf k})= -\frac{\eta_{\lambda\mu}({\bf
k})}{2bM^2\sigma_{\lambda\lambda}+\xi_0^2k^2}M. \eqno(15)
$$
Substitution of Eq. (15) in the definition of
$\sigma_{\lambda\lambda}$ renders the self-consistent equation:
$$
\sigma_{\lambda\lambda}=
\int\frac{<\eta_{\lambda\mu}(\bf{k})\eta_{\lambda\mu}(\bf{-k})>}
{[2bM^2\sigma_{\lambda\lambda}+\xi_0^2k^2]^2}\frac{d^3k}{(2\pi)^3},
\eqno(16)
$$
which has solution:
$$
\sigma_{\lambda\lambda}=\left(\frac{\Phi_{\lambda\mu\lambda\mu}(0)}
{8\pi\xi_0^3\sqrt{2|\tau|}}\right)^{2/3}; \eqno(17)
$$
and the same for
$$
\sigma_{\nu\nu}=\left(\frac{\Phi_{\nu\mu\nu\mu}(0)}
{8\pi\xi_0^3\sqrt{2|\tau|}}\right)^{2/3}. \eqno(17`)
$$
The correlation functions $\Phi_{\lambda\mu\lambda\mu}(0)$ and
$\Phi_{\nu\mu\nu\mu}(0)$ have the same order of magnitude as
$\Phi_{\mu\mu\mu\mu}(0)$. As a result, at $g\ll 1$ in the
mean-field approximation  the transverse fluctuations
$\sigma_{\lambda\lambda}$ and $\sigma_{\nu\nu}$ are of the order
$g^{2/3}$, i.e. more important then the longitudinal fluctuation
$\sigma_{\mu\mu}$. The integral in the self-consistent equation
(16) converges at $\xi_0^2k^2\sim bM^2\sigma_{\lambda\lambda}$, or
$k\sim(1/\xi(T))g^{1/3}$, i.e. at $g^{1/3}$ times smaller wave
vectors then those responsible for the longitudinal fluctuations.
The essential wave vectors are still much greater then
$1/L_{\eta}\sim g/\xi(T)$. This justifies the neglect in the above
argument of  possible variation of the orientation of $\bf{M}$
resulting from the Larkin-Imry-Ma effect. For $g\ll 1$ the
conditions of applicability of the perturbative approach
$\sigma_{\lambda\lambda}\ll 1$ and $\sigma_{\nu\nu}\ll 1$ are met.
Eq. (13)  can be considered as an extremum over $M^2$ of the
effective free energy density, which includes the principal
corrections due to the fluctuations:
$$
f_{eff}=\tau
M^2+\frac{1}{2}b(1+\sigma_{\lambda\lambda}+\sigma_{\nu\nu})M^4.
\eqno(18)
$$
The inclusion of the corrections results in the substitution of
$b(1+\sigma_{\lambda\lambda}+\sigma_{\nu\nu})$ instead of $b$. The
sum $\Sigma\equiv\sigma_{\lambda\lambda}+\sigma_{\nu\nu}$ may
depend on a direction of $\textbf{M}$. Another condition of
extremum of $f_{eff}$ is turning to zero  of the derivative of
$\Sigma$ over the angles, specifying orientation of
 $\bf{M}$. The energy gain at the transition is
$$
|\Delta
f_{eff}|=\frac{\tau^2}{2b(1+\sigma_{\lambda\lambda}+\sigma_{\nu\nu})}.
\eqno(19)
$$
Corrections $\sigma_{\lambda\lambda}$ and $\sigma_{\nu\nu}$ are
non-negative. The gain is maximum when both
$\sigma_{\lambda\lambda}=0$ and $\sigma_{\nu\nu}=0$. That raises a
question, for which ensemble $\eta_{jl}(\textbf{r})$ the maximum
gain can be reached. Eqns. (17) and (17`) show, that
$\sigma_{\lambda\lambda}$ and $\sigma_{\nu\nu}$ vanish if
$\eta_{\lambda\mu}=0$ and $\eta_{\nu\mu}=0$, i.e. if all
$\eta_{jl}(\textbf{r})$ have at least one principal direction in
common and magnetization is parallel to this direction:
$\eta_{jl}(\textbf{r})M_l=\varepsilon(\textbf{r}) M_j$. Here
$\varepsilon(\textbf{r})$ is a random eigenvalue. If all tensors
 $\eta_{jl}(\textbf{r})$ have three principal direction in common
the magnetization ${\bf M}$ can be oriented along either of these
directions.  According to Eqns.   (17), (17`), (19) the proper
orientation of ${\bf M}$ increases the energy gain for an amount
$\sim g^{2/3}(\tau^2/2b)$. This is much greater then the
 gain at the disruption of the long-range
order according to the Larkin-Imry-Ma mechanism, which is  $\sim
g^2(\tau^2/2b)$. The lifting of degeneracy suppresses the
Larkin-Imry-Ma mechanism  and the long-range order is preserved.

The described above lifting of continuous degeneracy by a random
perturbation is analogous to the lifting of degeneracy of the
mechanical pendulum with the vibrating suspension point (Kapitza
pendulum) \cite{kapitz}. If the gravity is disregarded the
equilibrium position of a pendulum is degenerate. The vibrations
lift the degeneracy and orient the pendulum so that there is no
torque in the direction of degeneracy of the pendulum.

For the magnet the perturbation has continuous spectrum of the
wave vectors and it is useful to elucidate the effect of different
parts of the spectrum on the order parameter. Small and large $k$
in comparison with $k_{\xi}\sim 1/\xi(T)$ have to be
distinguished. The short wavelength perturbations ($k\gg k_{\xi}$)
effect equally all degrees of freedom. The net effect of these
perturbations is a shift and, possibly, a splitting of the $T_c$,
which can be added to the corresponding terms in the free energy.
Difference between the longitudinal and transverse (Goldstone)
fluctuations occurs in the short wavelength region ($k\ll
k_{\xi}$). For the transverse degrees of freedom in the
three-dimensional case the main contribution to the free energy
comes from the fluctuations with $k\sim g^{1/3}k_{\xi}$. This
contribution is $g^{-1/3}$ times greater then for the longitudinal
degree of freedom. The disruption of the long-range order
according to the Larkin-Imry-Ma mechanism also comes via the
transverse fluctuations but with even smaller $k\sim gk_{\xi}$. If
the contribution of the transverse fluctuations to the free energy
depends on the orientation of the order parameter then the
minimization of this contribution favors the orientation of  ${\bf
M}$, which decreases the disrupting effect of the fluctuations on
the long-range order. If for some orientations of  ${\bf M}$ the
singular contribution of the Goldstone fluctuations can be
eliminated, then the Larkin-Imry-Ma mechanism of disruption of the
long-range order is fully suppressed.

{\bf 3.} Consider in more details the case of one common principal
direction assuming it as $z$-axis. The eigenvalues for that
direction $\eta_{33}(\textbf{r})$ are assumed to be statistically
independent with the other finite components of
 $\eta_{jl}(\textbf{r})$, which form a two-dimensional symmetric
 random tensor $\eta_{pq}(\textbf{r})$, where $p$ and $q$ take
 values 1 and 2. With regard to  the $\eta_{pq}(\textbf{r})$ it is
 assumed that they form isotropic ensemble in the $(xy)$-plane.
 Their correlation functions at $k\to 0$ can be expressed in terms
 of two constants:
$$
\Phi_{pqrs}(0)=<\eta_{pq}(\textbf{k})\eta_{rs}(\textbf{-k})>\mid_{k=0}=
\Phi^{(1)}\delta_{pq}\delta_{rs}+
\Phi^{(2)}(\delta_{pr}\delta_{qs}+\delta_{ps}\delta_{qr})
\eqno(20).
$$
The coefficients $\Phi^{(1)}$, $\Phi^{(2)}$ are related to the
averages
$\Phi^{(1)}=<\eta_{11}(\textbf{k})\eta_{22}(\textbf{-k})>\mid_{k=0},
\Phi^{(2)}=<\eta_{12}(\textbf{k})\eta_{12}(\textbf{-k})>\mid_{k=0},
\Phi^{(1)}+2\Phi^{(2)}=<\eta_{11}(\textbf{k})\eta_{11}(\textbf{-k})>\mid_{k=0}=
<\eta_{22}(\textbf{k})\eta_{22}(\textbf{-k})>\mid_{k=0}$. It
follows from these relations, that $\Phi^{(2)}>0$,
$\Phi^{(1)}+2\Phi^{(2)}>0$. For substitution in Eqns.  (18) and
(19) the correlation functions have to be expressed in the basis
$\hat{\lambda},\hat{\mu},\hat{\nu}$. The off-diagonal component
 $\sigma_{\lambda\nu}$ turns to zero if
 $\hat{\lambda}\parallel\hat{\mu}\times\hat{z}$ and
$\hat{\nu}=\hat{\lambda}\times\hat{\mu}$. Let $\theta$ be the
angle between  $\hat{\mu}$ and $z$-axis, then
$<\eta_{\lambda\mu}(\textbf{k})\eta_{\lambda\mu}(\textbf{-k})>\mid_{k=0}=
\Phi^{(2)}\sin^2\theta$,
$<\eta_{\nu\mu}(\textbf{k})\eta_{\nu\mu}(\textbf{-k})>\mid_{k=0}=
(\Phi^{(1)}+2\Phi^{(2)}+\Phi^{(3)})\sin^2\theta\cos^2\theta$,
where
$\Phi^{(3)}$=$<\eta_{\nu\mu}(\textbf{k})\eta_{\nu\mu}(\textbf{-k})>\mid_{k=0}$
$\Phi^{(3)}\geq 0$, for a two-dimensional disorder $\Phi^{(3)}=0$.
As a result,
$$
\Sigma=\left(\frac{1}
{8\pi\xi_0^3\sqrt{2|\tau|}}\right)^{2/3}[(\Phi^{(2)}\sin^2\theta)^{2/3}+
((\Phi^{(1)}+2\Phi^{(2)}+\Phi^{(3)})\sin^2\theta\cos^2\theta)^{2/3}]
  \eqno(21).
$$
$\Sigma=0$ at $\theta=0$ as it should be, but   $\theta=\pi/2$ is
also extremum.

This example makes possible to compare the orienting effect of
fluctuations with the average anisotropy, described by the term
$\kappa_{jl}M_jM_l$ in the free energy (1). Assume that anisotropy
is also uniaxial with the axis, parallel to  $\hat z$-direction,
then $\kappa_{jl}$ is specified by one parameter  $\kappa$:
$\kappa_{33}=2\kappa, \kappa_{11}=\kappa_{22}=-\kappa$. The trace
$\kappa_{ll}$ is included in the definition of $\tau$.  The
anisotropy splits $T_c$ in two -- for ${\bf M}$ in the plane
$(x,y)$ and perpendicular to it. The magnitude of splitting is
$\Delta\tau=3\kappa$. Depending on the sign of  $\kappa$ one or
the other orientation of ${\bf M}$ corresponds to higher $T_c$.
Since the random anisotropy always orients ${\bf M}$ along $\hat
z$ the case $\kappa>0$, when higher $T_c$ corresponds to the
orientation of ${\bf M}$ in the $(x,y)$-plane is more interesting.
With the account of the anisotropy
$$
f_{eff}=(\tau+2\kappa)M_3^2+(\tau-\kappa)(M_1^2+M_2^2)
+\frac{1}{2}b(1+\sigma_{\lambda\lambda}+\sigma_{\nu\nu})M^4.
\eqno(22)
$$
At $(\tau-\kappa)<0$ $f_{eff}$ has nontrivial extremum
$M_{\perp}^2\equiv(M_1^2+M_2^2)=-(\tau-\kappa)/b(1+\sigma_{\lambda\lambda}),
 M_3^2=0$. It corresponds to the state with the energy
$f_{\perp}=-(\tau-\kappa)^2/2b(1+\sigma_{\lambda\lambda})$. This
state is degenerate with respect to the direction of  ${\bf M}$ in
the $(x,y)$-plane and has to have according to Ref. \cite{feldm}
quasi-long-range order. At $(\tau+2\kappa)<0$ there appears
another extremum  $M_3^2=-(\tau+2\kappa)/b;
M_{\perp}^2\equiv(M_1^2+M_2^2)=0$. It corresponds to the state
with the long-range order and with the energy
$f_{\parallel}=-(\tau+2\kappa)^2/2b$. This state is more
advantageous at $\tau\approx -6\kappa/\sigma_{\lambda\lambda}$.
Transition between the two states according to Eq. (21) is of the
first order. The requirement of  applicability of the used
approximations imposes rather severe restriction on  the
anisotropy: $6\kappa/\sigma_{\lambda\lambda}\ll 1$.

Consider also the interplay of the disorder with the uniaxial
anisotropy in the fourth-order terms in  ${\bf M}$, when without
the disorder the free energy has a form:
$$
f=\tau M^2+\frac{1}{2}[b_{\parallel}M_3^4+b_{\perp}M_{\perp}^4].
\eqno(23)
$$
At $\tau<0$ the energy (23) has two extrema:
$M_3^2=-\tau/b_{\parallel}, M_{\perp}=0$ and $M_3^2=0,
M^2_{\perp}=-\tau/b_{\perp}$. The more advantageous is the state
with the smaller $b$. The disorder effects differently the two
states. More interesting is the case $b_{\perp}<b_{\parallel}$,
when without the disorder orientation of ${\bf M}$  in the
$(x,y)$-plane is favored. The inclusion of the principal order
corrections due to the random anisotropy amounts to the
substitution of $b_{\perp}(1+\sigma_{\lambda\lambda})$ instead of
 $b_{\perp}$ in the solution. Another solution  ${\bf
M}\parallel\hat{z}$ in the same approximation does not have
corrections and for
$\sigma_{\lambda\lambda}>(b_{\parallel}-b_{\perp})/b_{\perp}$ it
became more advantageous. In the temperature interval  from
 $\tau=0$ till the temperature when
$\sigma_{\lambda\lambda}=(b_{\parallel}-b_{\perp})/b_{\perp}$ the
disorder interchanges two states. In the vicinity of $T_c$ the
state with the long-range order is stabilized.  Since it has been
assumed, that $\sigma_{\lambda\lambda}\ll 1$, the interchange is
possible only for the small anisotropy, i.e. when
$(b_{\parallel}-b_{\perp})/b_{\perp}\ll 1$.

I grateful to  A.F. Andreev, E.I. Kats and V.I. Marchenko for the
useful discussions.  This work is partly supported by RFBR (grant
07-02-00214), Ministry of Science and Education of the Russian
Federation and CRDF (grant RUP1-2632-MO04).


\begin{thebibliography}{99}


\bibitem{Lark} A. I. Larkin, Zh. Exp. Teor. Fiz., {\bf 58}, 1466 (1970) [Sov. Phys. JETP,
{\bf 31}, 784 (1970)]

\bibitem{imry} Y. Imry and S. Ma,  Phys. Rev. Lett.
               {\bf 35}, 1399 (1975)

\bibitem{aharon} A. Aharony,  Phys. Rev.
               {\bf B18}, 3328 (1978)

\bibitem{minchau} B. J. Minchau and R. A. Pelcovits,  Phys. Rev.
               {\bf B32}, 3081 (1985)

\bibitem{wehr} J. Wehr, A. Niederberger, L. Sanches-Palencia and M. Lewenstein
                Phys. Rev. {\bf B74}, 224448 (2006)

\bibitem{fom1} I. A. Fomin, Zh. Exp. Teor. Fiz. {\bf 125}, 1115 (2004)
[JETP, {\bf 98}, 974 (2004)]

\bibitem{plish} R. Harris, M. Plishke, and M. J. Zuckermann, Phys. Rev. Lett.
               {\bf 31}, 160 (1973)

\bibitem{LarkOv}  A.I.Larkin and Yu.N. Ovchinnikov, Zh. Exp. Teor. Fiz. {\bf 61}, 1221
(1971),[Sov. Phys. JETP, {\bf 34}, 651 (1971)]

\bibitem{kapitz} P. L. Kapitza, Zh. Exp. Teor. Fiz. {\bf 21}, 588 (1951).

\bibitem{feldm} D. E. Feldman, Int. Journ. Mod. Phys, {\bf 15}, 2954 (2001).






\end{thebibliography}
\end{document}